\documentclass[aps,prd,nofootinbib,preprint,floatfix]{revtex4}
\usepackage{amsfonts}
\usepackage{amsmath}
\usepackage{hyperref}
\usepackage{amssymb}
\usepackage[english]{babel}
\usepackage{graphicx}
\usepackage{epsfig}
\usepackage{bm}
\usepackage{verbatim}
\usepackage[utf8]{inputenc}
\usepackage{color}
\usepackage{xcolor}
\usepackage{subfig}
\usepackage[normalem]{ulem}
\usepackage{dsfont}
\usepackage{multirow}

\newcommand{\mathsym}[1]{{}}

\newcommand{\emp}{\begin{equation}}
\newcommand{\fin}{\end{equation}}
\newcommand{\empn}{\begin{equation*}}
\newcommand{\finn}{\end{equation*}}

\newcommand{\bea}{\begin{eqnarray}}
\newcommand{\eea}{\end{eqnarray}}
\newcommand{\eger}{\begin{gather}}
\newcommand{\fger}{\end{gather}}
\newcommand{\egn}{\begin{gather*}}
\newcommand{\fgn}{\end{gather*}}
\newcommand{\bit}{\begin{itemize}}
\newcommand{\eit}{\end{itemize}}

\begin{document}

\title{Low scale type I seesaw model for lepton masses and mixings}
\author{A. E. C\'arcamo Hern\'andez}
\email{antonio.carcamo@usm.cl}
\affiliation{{\small Universidad T\'ecnica Federico Santa Mar\'{\i}a and Centro Cient\'{\i}fico-Tecnol\'ogico de Valpara\'{\i}so}\\
Casilla 110-V, Valpara\'{\i}so, Chile}
\author{Marcela Gonz\'alez}
\email{marcela.gonzalezp@usm.cl}
\affiliation{{\small Universidad T\'ecnica Federico Santa Mar\'{\i}a and Centro Cient\'{\i}fico-Tecnol\'ogico de Valpara\'{\i}so}\\
Casilla 110-V, Valpara\'{\i}so, Chile}
\author{Nicol\'as A. Neill}
\email{nicolas.neill@gmail.com}
\affiliation{{\small Universidad T\'ecnica Federico Santa Mar\'{\i}a and Centro Cient\'{\i}fico-Tecnol\'ogico de Valpara\'{\i}so}\\
Casilla 110-V, Valpara\'{\i}so, Chile}
\date{\today }

\begin{abstract}
In contrast to the original type I seesaw mechanism that requires
right-handed Majorana neutrinos at energies much higher than the electroweak
scale, the so-called low scale seesaw models allow lighter masses for the
additional neutrinos. Here we propose an alternative low scale type I seesaw
model, where neither linear nor inverse seesaw mechanisms take place, but
the spontaneous breaking of a discrete symmetry at an energy scale much
lower than the model cutoff is responsible for the smallness of the light
active neutrino masses. In this scenario, the model is defined with minimal
particle content, where the right-handed Majorana neutrinos can have masses
at the $\sim 50\mbox{ GeV}$ scale. The model is predictive in the neutrino
sector having only four effective parameters that allow to successfully
reproduce the experimental values of the six low energy neutrino observables.
\end{abstract}

\maketitle

\section{Introduction}

After minimally extending the Standard Model (SM) to include massive
neutrinos, the observed fermion mass hierarchy is extended over a range of $13$ orders of magnitude, from the lightest active neutrino mass scale up to
the top quark mass. In addition, the small quark mixing angles decrease from
one generation to the next while in the lepton sector this hierarchy is not
present since two of the mixing angles are large and the other one is small.
Neither of these features in the flavor sector is explained in the SM. This
is the so-called \textit{SM flavor puzzle}, which has motivated the
construction of theories with extended scalar and/or fermion sectors with
additional continuous or discrete groups. In particular, extensions of the
SM with non-Abelian discrete flavor symmetries are very attractive since
they successfully describe the observed pattern of fermion masses and
mixings (for recent reviews on discrete flavor groups see Refs.~\cite%
{Ishimori:2010au,Altarelli:2010gt,King:2013eh,King:2014nza,King:2017guk,Petcov:2017ggy}%
), while they can naturally appear from the breaking of continuous
non-Abelian gauge symmetries or from compactified extra dimensions (see Ref.~%
\cite{King:2019gif} and references therein).
Several discrete groups have been employed in extensions of the SM.
In particular, $A_{4}$ is the smallest discrete group with one three-dimensional and three distinct one-dimensional irreducible representations where the three families of fermions can be accommodated rather naturally.
This group has been particularly promising in providing a predictive description of the current pattern of SM fermion masses and mixing angles \cite%
{Ma:2001dn,He:2006dk,Chen:2009um,Feruglio:2009hu,Altarelli:2012bn,Ahn:2012tv,Memenga:2013vc,Morisi:2013eca,Felipe:2013vwa,Varzielas:2012ai, Ishimori:2012fg,King:2013hj,Hernandez:2013dta,Babu:2002dz,Altarelli:2005yx,Gupta:2011ct,Morisi:2013eca, Altarelli:2005yp,Kadosh:2010rm,Kadosh:2013nra,delAguila:2010vg,Campos:2014lla,Vien:2014pta,Joshipura:2015dsa,Hernandez:2015tna,Karmakar:2016cvb,Borah:2017dmk,Chattopadhyay:2017zvs,CarcamoHernandez:2017kra,Ma:2017moj,CentellesChulia:2017koy,Bjorkeroth:2017tsz,Srivastava:2017sno,Borah:2018nvu,Belyaev:2018vkl,CarcamoHernandez:2018aon,Srivastava:2018ser,delaVega:2018cnx,Pramanick:2019qpg,CarcamoHernandez:2019pmy,Okada:2019uoy}%
.
Despite several models based on the $A_{4}$ discrete symmetry have been
proposed, most of them have a nonminimal scalar sector, 
composed of several $SU(2)$ Higgs doublets, even in their low energy limit,
and have $A_{4}$ scalar triplets in the scalar spectrum whose vacuum
expectation value (VEV) configurations in the $A_{4}$ direction are not the
most natural solutions of the scalar potential minimization equations.
Thus, it would be desirable to build an $A_{4}$ flavor model which at low energies reduces to the SM model and where the different gauge singlet scalars are accommodated into $A_{4}$ singlets and one $A_{4}$ triplet [with VEV pattern
in the $(1,1,1)$ $A_{4}$ direction] which satisfies the minimization condition of the scalar potential for the whole range of values of the parameter space.
To this end, in this work we propose an extension of the SM based on the $A_{4}$ family symmetry, which
is supplemented by a $Z_4$ auxiliary symmetry, whose spontaneous breaking at
an energy scale ($v_S$) much lower than the model cutoff ($\Lambda$)
produces the small light active neutrino mass scale $m_\nu$. As we will show
in the next sections, in this scenario the masses for the active neutrinos
are produced by a type I seesaw mechanism \cite%
{Schechter:1980gr,Schechter:1981cv,King:1999mb,King:2002nf} mediated by
three $\sim 50\mbox{ GeV}$ right-handed Majorana neutrinos, where $%
m_\nu\propto \left(v_S/\Lambda \right)^2$. Given the low mass scale of the
right-handed neutrinos, this model can be classified as a \textit{low scale}
type I seesaw, as it has been coined in the literature \cite%
{Gluza:2002vs,Kersten:2007vk,Das:2017nvm,Xing:2009in,He:2009ua,Adhikari:2010yt,Ibarra:2010xw,Ibarra:2011xn,Lopez-Pavon:2015cga,Cely:2012bz,Dev:2013oxa}%
. There are different realizations of low scale seesaw models, as for
example \textit{inverse} or \textit{linear} \cite%
{CarcamoHernandez:2017kra,CarcamoHernandez:2019pmy,Mohapatra:1986bd,GonzalezGarcia:1988rw,Akhmedov:1995vm,Akhmedov:1995ip,Malinsky:2005bi,Dev:2009aw,Abada:2014vea,Dib:2014fua,CarcamoHernandez:2017cwi,CarcamoHernandez:2017owh,CarcamoHernandez:2018iel,CarcamoHernandez:2018hst,CarcamoHernandez:2019eme,CarcamoHernandez:2019vih,Dib:2019jod,CarcamoHernandez:2020pnh}
, where an additional lepton number violating mass parameter is added. In
these models, the smallness of $m_\nu$ is related to the smallness of the
additional parameter. In our case, however, no extra small mass parameter
has been included, and the smallness of the light neutrino masses is explained through the spontaneous breaking of the auxiliary
discrete groups, which leads to a suppression in the Dirac neutrino mass
matrix.

From the point of view of the low energy neutrino observables,
the model makes very particular predictions for $\delta_{CP}$ and $%
\theta_{23}$, which are not aligned with the central values of current fits.
Therefore, future improvements in the precision of neutrino measurements will provide an experimental test of the model. Processes like (i) charged
lepton flavor violating decays ($\ell\to \ell^{\prime }\gamma$) \cite%
{Feruglio:2009hu,Altarelli:2012bn}, (ii) flavor changing neutral currents,
and (iii) rare top quark decays such as $t\to hc$, $t\to cZ$ \cite%
{Morisi:2013eca}, are strongly suppressed, in contrast to other $A_4$ flavor
models (that usually have several Higgs doublets), where these processes can
have rates that are at the reach of forthcoming experiments.

The paper is organized as follows. In Sec.~\ref{model} we describe the
model. In Sec.~\ref{leptonsector} we present a discussion on lepton
masses and mixings and give the corresponding results. We draw our
conclusions in Sec.~\ref{conclusions}. The Appendix provides a
concise description of the $A_{4} $ discrete group.

\section{Model Description}

\label{model}

We propose an extension of the SM where the scalar sector is augmented by
the inclusion of four gauge-singlet scalar fields and the SM gauge symmetry
is supplemented by the $A_{4}\times Z_{4}$ discrete group. The symmetry $%
\mathcal{G}$ features the following spontaneous symmetry breaking pattern: 
\begin{eqnarray}
\mathcal{G}= SU\left( 2\right)_L \times U\left( 1\right)_{Y}\times
A_{4}\times Z_{4} \xrightarrow{v_S}, SU\left( 2\right) _{L}\times U\left(
1\right)_{Y}{\xrightarrow{v}}U\left( 1 \right)_{Q},
\end{eqnarray}

where the symmetry-breaking scales satisfy the hierarchy $v_S \sim \mathcal{O%
}(1)\mbox{TeV} >v$, $v_S$ is the scale of spontaneous breaking of the $%
A_{4}\times Z_{4}$ discrete group, and $v=246\mbox{ GeV}$ is the electroweak
symmetry breaking scale. As mentioned before, the scalar sector of the SM is
augmented by the inclusion of four SM gauge singlet scalars. We add these
extra scalar fields for the following reasons: (i) to build
nonrenormalizable charged leptons and Dirac neutrino Yukawa terms invariant
under the local and discrete groups, crucial to generate predictive textures
for the lepton sector; (ii) to generate a renormalizable Yukawa term for
the right-handed Majorana neutrinos, that can give rise to $\sim 50\mbox{ GeV}
$ masses for these singlet fermions. As we will see below, the observed
pattern of SM charged lepton masses and leptonic mixing angles will arise
from the spontaneous breaking of the $A_{4}\times Z_{4}$ discrete group. In
order to generate the masses for the light active neutrinos via a type-I
seesaw mechanism, we extend the fermion sector by including three
right-handed Majorana neutrinos, which are singlets under the SM group. The
lepton assignments under the group $A_{4}\times Z_{4}$ are 
\begin{eqnarray}
L_{L} &\sim &\left( \mathbf{3},0\right) ,\hspace{1.5cm}N_{R}\sim \left( 
\mathbf{3},1\right) ,  \notag \\
l_{1R} &\sim &\left(\mathbf{1}^{\prime}\mathbf{,}1\right),\hspace{ 1.5cm}%
l_{2R}\sim \left( \mathbf{1,}1\right) ,\hspace{1.5cm}l_{3R}\sim \left( 
\mathbf{\mathbf{1}^{\prime\prime }},1\right) .  \label{fermionassignments}
\end{eqnarray}
Here we specify the dimensions of the $A_{4}$ irreducible representations by
the numbers in boldface and we write the different $Z_{4}$ charges in
additive notation. Regarding the lepton sector, note that the left- and
right-handed leptonic fields are accommodated into $A_{4}$ triplet and $A_{4}
$ singlet irreducible representations, respectively, whereas the
right-handed Majorana neutrinos are unified into an $A_{4}$ triplet. The
scalar spectrum of the model includes the SM Higgs doublet $\phi$ and the
gauge singlet scalars $\rho$, $\xi _{j}$ ($j=1,2,3$). The scalar fields have
the following transformation properties under the flavor symmetry $%
A_{4}\times Z_{4}$: 
\begin{eqnarray}
\phi &\sim &\left( \mathbf{1,}0\right) ,\hspace{1.5cm}\rho _{1}\sim \left( 
\mathbf{1}^{\prime \prime }\mathbf{,-}1\right) ,\hspace{1.5cm}\rho _{2}\sim
\left( \mathbf{\mathbf{1}^{\prime },-}1\right) ,\hspace{1.5cm}  \notag \\
\rho _{3} &\sim &\left( \mathbf{\mathbf{1},-}2\right) ,\hspace{1.5cm}\xi
\sim \left( \mathbf{3,}-1\right) .
\end{eqnarray}
We assume the following vacuum configuration for the $A_{4}$-triplet gauge
singlet scalar $\xi$: 
\begin{equation}
\left\langle \xi \right\rangle =\frac{v_{\xi }}{\sqrt{3}}\left( 1,1,1\right)
,  \label{VEV}
\end{equation}
which satisfies the minimization condition of the scalar
potential for the whole range of values of the parameter space, as shown in
Ref. \cite{Hernandez:2015tna}. With the particle content previously
specified, we have the following relevant Yukawa terms for the lepton
sector, invariant under the symmetries of the model: 
\begin{eqnarray}
&\mathcal{L} _{Y}^{\left( L\right) } = &y_{1}^{\left( L\right) }\left(%
\overline{L}_{L}\phi \xi \right) _{\mathbf{1^{\prime \prime }}} l_{1R} \frac{%
1}{\Lambda} +y_{2}^{\left( L\right) }\left( \overline{L}_L \phi \xi \right)_{%
\mathbf{1}} l_{2R} \frac{1}{\Lambda } +y_{3}^{\left( L\right) }\left( 
\overline{L}_L \phi \xi \right)_{\mathbf{1^{\prime }}} l_{3R}\frac{1}{\Lambda%
}  \notag \\
&& + y_{N}\left( \overline{N}_R N_R^C \right)_{\mathbf{1}}
(\rho_{3}+c\rho_{3}^*) + y_{1\nu} \left(\overline{L}_L \widetilde{\phi}
N_{R} \right)_{\mathbf{1^{\prime }}} \frac{\rho_1}{\Lambda} + y_{2\nu}\left(%
\overline{L}_L \widetilde{\phi } N_{R} \right)_{\mathbf{1^{\prime \prime }}}%
\frac{\rho_2}{\Lambda}  \notag \\
&& + y_{3\nu }\left( \overline{L}_L \widetilde{\phi }N_{R}\right)_{\mathbf{3s%
}} \frac{\xi}{\Lambda } + y_{4\nu }\left( \overline{L}_{L}\widetilde{\phi }
N_{R}\right) _{\mathbf{3a}}\frac{\xi }{\Lambda } +\mbox{h.c.},  \label{lyl}
\end{eqnarray}
where the dimensionless couplings in Eq. (\ref{lyl}) are $\mathcal{O}(1)$
parameters.

In what follows, we describe the role of each discrete group factor of our $%
A_{4}$ flavor model. The $A_{4}$ discrete group yields a reduction of the
number of model parameters, giving rise to predictive textures for the
lepton sector, which are consistent with the lepton mass and mixing pattern,
as will be shown in Sec.~\ref{leptonsector}. On the other hand, the $Z_{4}
$ discrete group is the smallest cyclic symmetry allowing a
renormalizable Yukawa term for the right-handed Majorana neutrinos, giving
rise to a diagonal Majorana neutrino mass matrix that yields degenerate
Majorana neutrinos with electroweak scale masses. In addition, the
spontaneous breaking of the $A_4\times Z_{4}$ discrete group at an energy
scale much lower than the model cutoff is crucial to produce small mixing
mass terms between the active and sterile neutrinos, allowing the
implementation of a low scale type I seesaw mechanism. Finally, we assume
that the VEVs of the gauge singlet scalar fields $\xi$, $\rho_{i}$ ($i=1,2,3$%
) satisfy the relation
\begin{equation}
v_{\rho_j} \sim v_\xi \sim \mathcal{O}(1)\mbox{TeV} \ll \Lambda,\hspace{1cm}%
j=1,2,3,
\end{equation}
where $v_\xi \sim v_\rho \sim v_S$ is the discrete symmetry breaking scale
and $\Lambda$ is the model cutoff.

It is worth mentioning that this model at low energies corresponds to a
singlet-doublet model \cite{Schabinger:2005ei,Patt:2006fw}. Consequently,
from a detailed analysis of the low energy scalar potential (as done for
example in Ref. \cite{CarcamoHernandez:2019cbd}) one can show that the $125%
\mbox{ GeV}$ SM-like Higgs boson has couplings close to the SM expectation,
with small deviations of order $v^2/v_S^2 \sim \mathcal{O}(10^{-2})$.
The TeV-scale singlet $s^0$ ($s^0=\xi,\rho_j$) will mix with the $CP$-even
neutral component of the SM Higgs doublet, $h^0$, with a mixing angle $%
\gamma \sim \mathcal{O}(v/v_S)$. Thus, the couplings of the singlet scalars
to the SM particles will be equal to the SM Higgs couplings times the $%
s^0-h^0$ mixing angle $\gamma$. The collider phenomenology of this scenario
is well studied \cite%
{Bertolini:2012gu,Robens:2015gla,Falkowski:2015iwa,Gorbahn:2015gxa,Buttazzo:2015bka}%
. For TeV-scale singlets, the most stringent limits at the 8TeV LHC come
from indirect searches. A global fit to all SM signal strengths constrains $%
\sin^2\gamma \leq 0.23$ at 95\% C.L. \cite%
{Giardino:2013bma,Falkowski:2013dza}, that assuming $\mathcal{O}(1)$
couplings in the scalar potential translates to $v_S\gtrsim 500\mbox{ GeV}$.
For a summary of the sensitivity of future colliders see for example Table 1
of Ref. \cite{Buttazzo:2015bka}. As we will see in the next section, there
is a broad range of values of $v_S$ that are consistent with the observed
light neutrino masses and current limits on singlet scalars.

\section{Neutrino masses and mixings}

\label{leptonsector}

The lepton Yukawa terms in Eq. (\ref{lyl}) imply that the mass matrix for
charged leptons is given by
\begin{equation}
M_{l}=V_{lL}S_{\ell }~\mbox{diag}\left( m_{e},-m_{\mu },m_{\tau }\right),
\end{equation}
where 
\begin{equation}
\hspace{1cm} V_{lL}=\frac{1}{\sqrt{3}}\left( 
\begin{array}{ccc}
1 & 1 & 1 \\ 
1 & \omega ^{2} & \omega \\ 
1 & \omega & \omega ^{2}%
\end{array}
\right) ,\hspace{1cm}S_{\ell }=\left( 
\begin{array}{ccc}
0 & -1 & 0 \\ 
1 & 0 & 0 \\ 
0 & 0 & 1%
\end{array}%
\right) ,\hspace{1cm}\omega =e^{\frac{2\pi i}{3}},  \label{Ml}
\end{equation}
so, 
\begin{equation}
M_{l}= \frac{1}{\sqrt{3}}\allowbreak \left( 
\begin{array}{ccc}
m_{e} & m_{\mu } & m_{\tau } \\ 
\omega ^{2} m_{e} & m_{\mu } & \omega m_{\tau } \\ 
\omega m_{e} & m_{\mu } & \omega^{2} m_{\tau }%
\end{array}
\right).
\end{equation}

Regarding the neutrino sector, we find that the resulting Dirac neutrino
mass matrix reads
\begin{equation}
M_{\nu }^{D}=\left( 
\begin{array}{ccc}
b+c & a+d & a-d \\ 
a-d & \omega b+c\omega ^{2} & a+d \\ 
a+d & a-d & \omega ^{2}b+c\omega%
\end{array}
\right) \frac{vv_S}{\sqrt{2}\Lambda},
\end{equation}
where $\omega =e^{\frac{2\pi i}{3}}$ and $a$, $b$, $c$, $d$ are effective
parameters related to the neutrino Yukawa couplings in Eq.~(\ref{lyl}).

The fermion sector is extended by including three right-handed Majorana neutrinos with masses $m_N$, where the Majorana mass matrix $M_N$ is proportional to the identity, $M_N=m_N ~ \hat{\mathbf{1}}_{3\times 3}$.
Given that $m_{N}\gg \left( M_{\nu }^{D}\right) _{ij}$ ($i,j=1,2,3$), the
light active neutrino mass matrix ($M_{\nu }$) arises from a type I seesaw
mechanism: 
\begin{eqnarray}
M_{\nu } &=&M_{\nu }^{D}M_{N}^{-1}\left( M_{\nu }^{D}\right) ^{T} \\
&=&\frac{1}{m_{N}}\left( 
\begin{array}{ccc}
b+c & a+d & a-d \\ 
a-d & \omega b+c\omega ^{2} & a+d \\ 
a+d & a-d & \omega ^{2}b+c\omega 
\end{array}%
\right) \left( 
\begin{array}{ccc}
b+c & a-d & a+d \\ 
a+d & c\omega ^{2}+b\omega  & a-d \\ 
a-d & a+d & b\omega ^{2}+c\omega 
\end{array}%
\right) \frac{v^{2}v_{S}^{2}}{2\Lambda ^{2}},\label{eq:mnu}
\end{eqnarray}%
where we can read that the typical mass scale of the light active neutrinos is 
\begin{eqnarray}
m_\nu \sim \frac{v^2}{2 m_N}\left(\frac{v_S}{\Lambda} \right)^2.\label{mnu}
\end{eqnarray}
It is noteworthy that the smallness of the active neutrino masses is a
consequence of their inverse scaling with the square of the model
cutoff, which is much larger than the breaking scale ($v_{S}$) of the
discrete symmetries. We can see from Eq.~(\ref{mnu}) that for heavy
neutrinos with masses $m_{N}\sim \mathcal{O}(50\mbox{ GeV})$, there is a
wide range of values of $v_{S}$ that produce the required suppression,
depending on the specific value of the model cutoff.
To show that the model is consistent with the neutrino oscillation
experimental data, we fix $m_{\nu }=50\mbox{ meV}$ and vary the neutrino
sector parameters $a$, $b$, $c$ and $d$, to adjust the neutrino mass squared
splittings $\Delta m_{21}^{2}$, $\Delta m_{31}^{2}$, the leptonic mixing
angles $\theta _{12}$, $\theta _{13}$, $\theta _{23}$, and the leptonic Dirac
CP violating phase $\delta _{CP}$ to their experimental values.
\begin{table}[t]
\footnotesize
\begin{center}
\renewcommand{\arraystretch}{1} 
\begin{tabular}{c|c||c|c|c|c}
\hline
\multirow{2}{*}{Observable}&\multirow{2}{*}{\parbox{7em}{Average\\ model value}  }
&\multicolumn{4}{|c}{Neutrino oscillation global fit values (NH)}\\
\cline{3-6}
 & & Best fit $\pm 1\sigma$ \cite{deSalas:2017kay} & Best fit $%
\pm 1\sigma$ \cite{Esteban:2018azc} & $3\sigma$ range \cite{deSalas:2017kay}
& $3\sigma$ range \cite{Esteban:2018azc} \\ \hline\hline
$\Delta m_{21}^{2}$ [$10^{-5}$eV$^{2}$] & $7.61$ & $%
7.55_{-0.16}^{+0.20}$ & $7.39_{-0.20}^{+0.21}$ & $7.05-8.14$ &  $6.79-8.01$ \\ \hline
$\Delta m_{31}^{2}$ [$10^{-3}$eV$^{2}$] & $2.51$ & $2.50\pm 0.03$
& $2.525_{-0.032}^{+0.033}$ & $2.41-2.60$ & $2.427-2.625$
\\ \hline
$\theta _{12}(^{\circ })$ & $34.1$ & $34.5_{-1.0}^{+1.2}$ &  $33.82_{-0.76}^{+0.78}$ & $31.5-38.0$ & $31.61-36.27$ \\ 
\hline
$\theta _{13}(^{\circ })$ & $8.45$ & $8.45_{-0.14}^{+0.16}$ &  $8.61\pm 0.13$ & $8.0-8.9$ & $8.22-8.99$ \\ \hline
$\theta _{23}(^{\circ })$ & $42.8$ & $47.7_{-1.7}^{+1.2}$ &  $49.6_{-1.2}^{+1.0}$ & $41.8-50.7$ & $40.3-52.4$ \\ \hline
$\delta _{CP}(^{\circ })$ & $313$ & $218_{-27}^{+38}$ & $%
215_{-29}^{+40}$ & $157-349$ & $125-392$ \\ \hline\hline
\end{tabular}%
\normalsize
\end{center}
\caption{\emph{Normal mass hierarchy}.---Model and experimental values of the
neutrino mass squared splittings, leptonic mixing angles, and $CP$-violating
phase.
The second column shows the average model value for each observable, calculated from the model solutions that reproduce the neutrino observables at the 90\% C.L.
The experimental values are taken from Refs.~\protect\cite%
{deSalas:2017kay,Esteban:2018azc}.}
\label{tab:neutrinos-NH}
\end{table}
\begin{table}[h]
\footnotesize
\begin{center}
\renewcommand{\arraystretch}{1} 
\begin{tabular}{c|c||c|c|c|c}
\hline
\multirow{2}{*}{Observable}&\multirow{2}{*}{\parbox{7em}{Average\\ model value}  }
&\multicolumn{4}{|c}{Neutrino oscillation global fit values (IH)}\\
\cline{3-6}
 & & Best fit $\pm 1\sigma$ \cite{deSalas:2017kay} & Best fit $%
\pm 1\sigma$ \cite{Esteban:2018azc} & $3\sigma$ range \cite{deSalas:2017kay}
& $3\sigma$ range \cite{Esteban:2018azc} \\ \hline\hline
$\Delta m_{21}^{2}$ [$10^{-5}$eV$^{2}$] & $7.61$ & $%
7.55_{-0.16}^{+0.20}$ & $7.39_{-0.20}^{+0.21}$ & $7.05-8.14$ & 
 $6.79-8.01$ \\ \hline
$\Delta m_{13}^{2}$ [$10^{-3}$eV$^{2}$] & $2.41$ & $%
2.42_{-0.04}^{+0.03}$ & $2.512_{-0.032}^{+0.034}$ & $2.31-2.51$
& $2.412-2.611$ \\ \hline
$\theta_{12} (^{\circ })$ & $34.7$ & $34.5_{-1.0}^{+1.2}$
& $33.82_{-0.76}^{+0.78}$ & $31.5-38.0$ & $31.61-36.27$ \\ 
\hline
$\theta_{13} (^{\circ })$ & $8.56$ & $8.53_{-0.15}^{+0.14}$
& $8.65\pm 0.13$ & $8.1-9.0$ & $8.27-9.03$ \\ \hline
$\theta_{23} (^{\circ })$ & $48.7$ & $47.9_{-1.7}^{+1.0}$
& $49.8_{-1.1}^{+1.0}$ & $42.3-50.7$ & $40.6-52.5$ \\ 
\hline
$\delta_{CP} (^{\circ })$ & $297$ & $281_{-27}^{+23}$ &  $284_{-29}^{+27}$ & $202-349$ & $196-360$ \\ \hline
\end{tabular}%
\end{center}
\caption{\emph{Inverted mass hierarchy}.---Model and experimental values of the
neutrino mass squared splittings, leptonic mixing angles, and $CP$-violating
phase.
The second column shows the average model value for each observable, calculated from the model solutions that reproduce the neutrino observables at the 90\% C.L.
The experimental values are taken from Refs.~\protect\cite%
{deSalas:2017kay,Esteban:2018azc}.}
\label{tab:neutrinos-IH}
\normalsize
\end{table}
\begin{figure}[t]
\subfloat
{\ \includegraphics[width=0.45\textwidth]{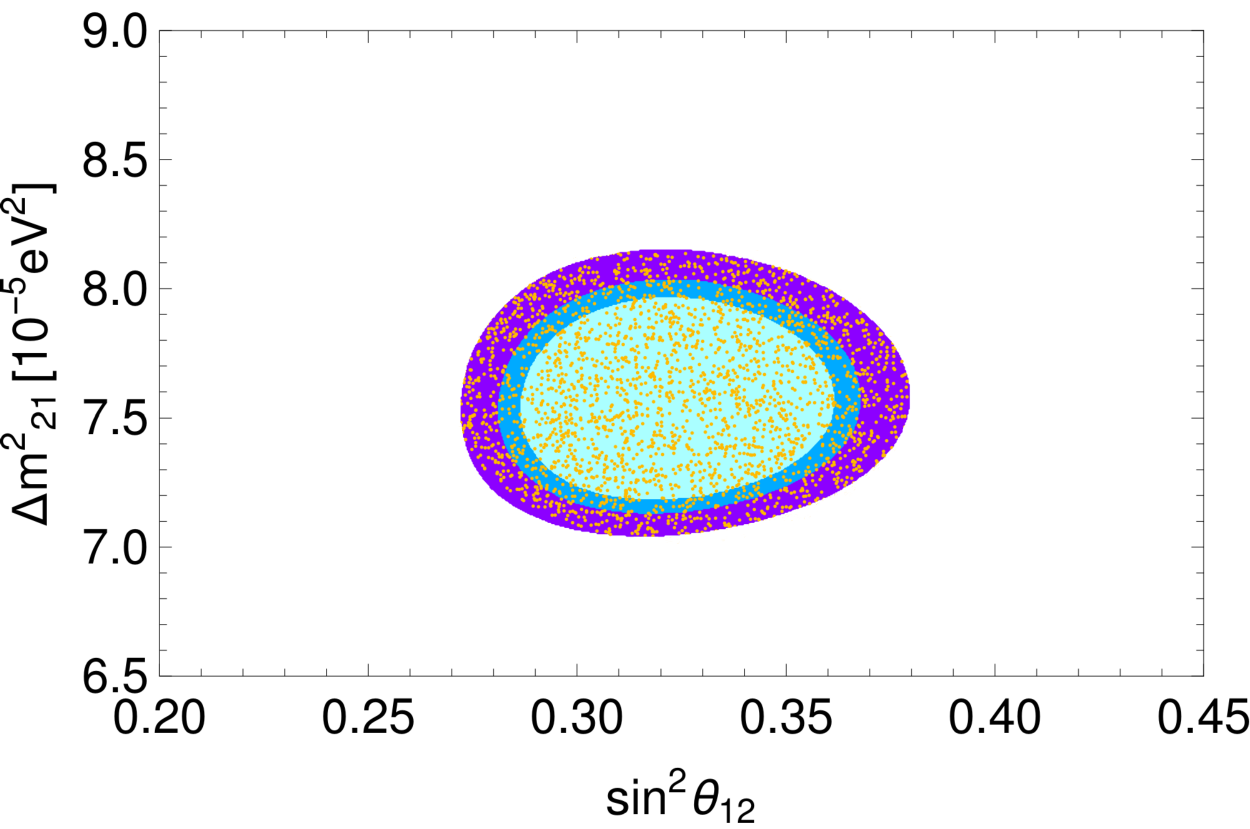}  } \subfloat
{\ \includegraphics[width=0.437\textwidth]{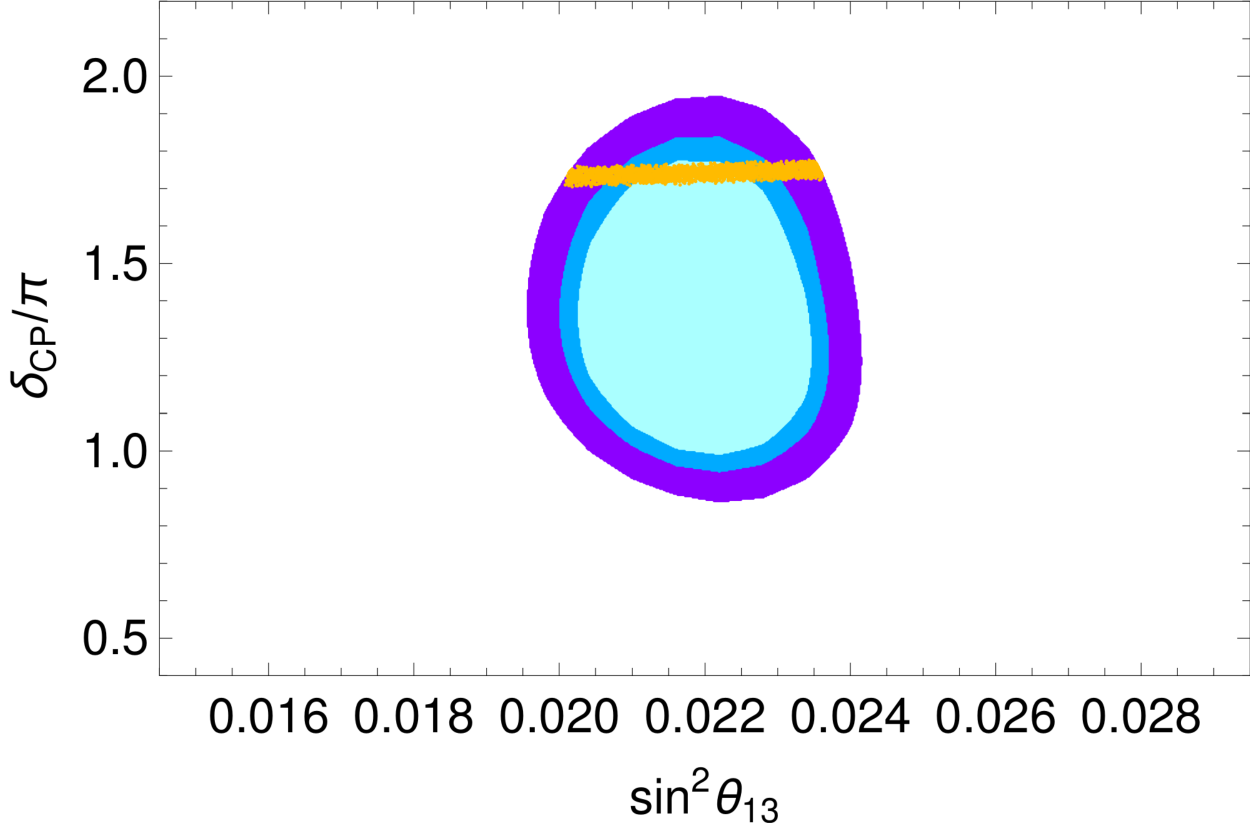}  } \hfill \subfloat
{\ \includegraphics[width=0.45\textwidth]{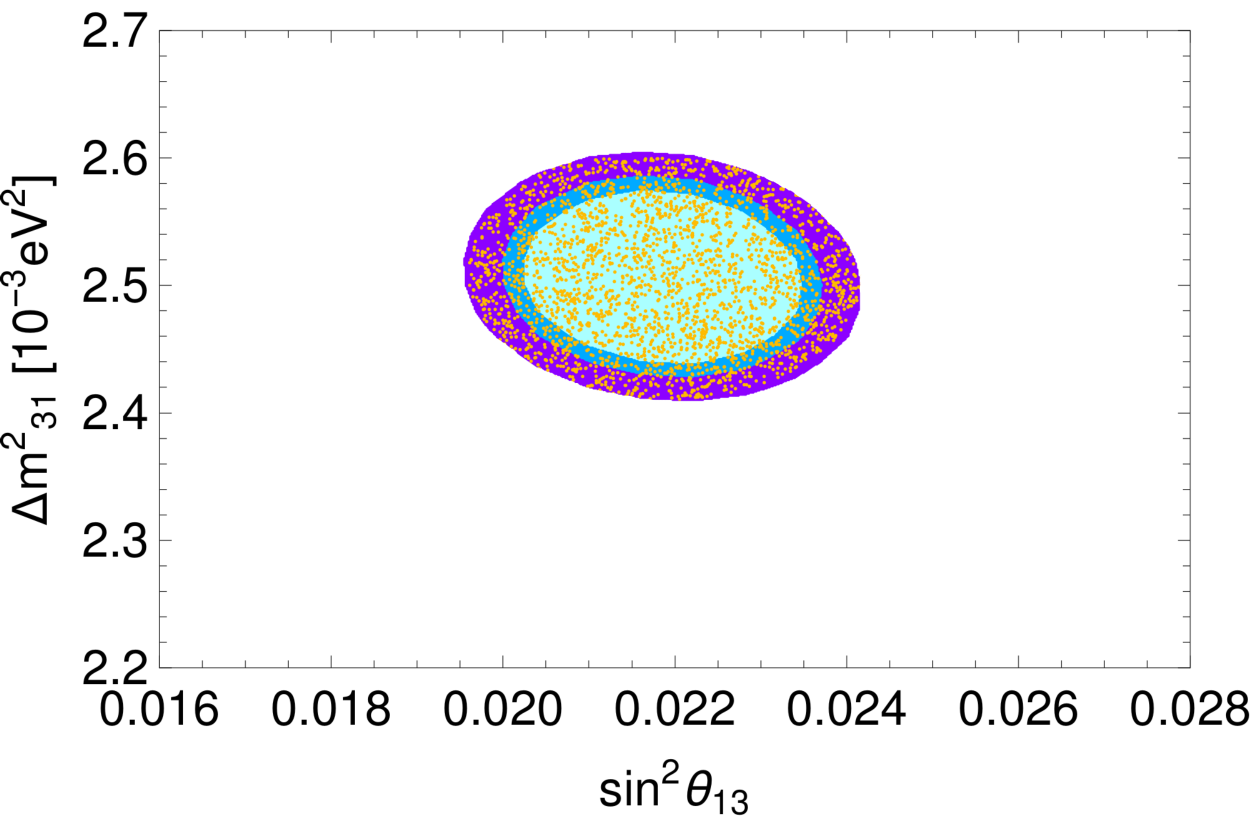}  } \subfloat
{\ \includegraphics[width=0.432\textwidth]{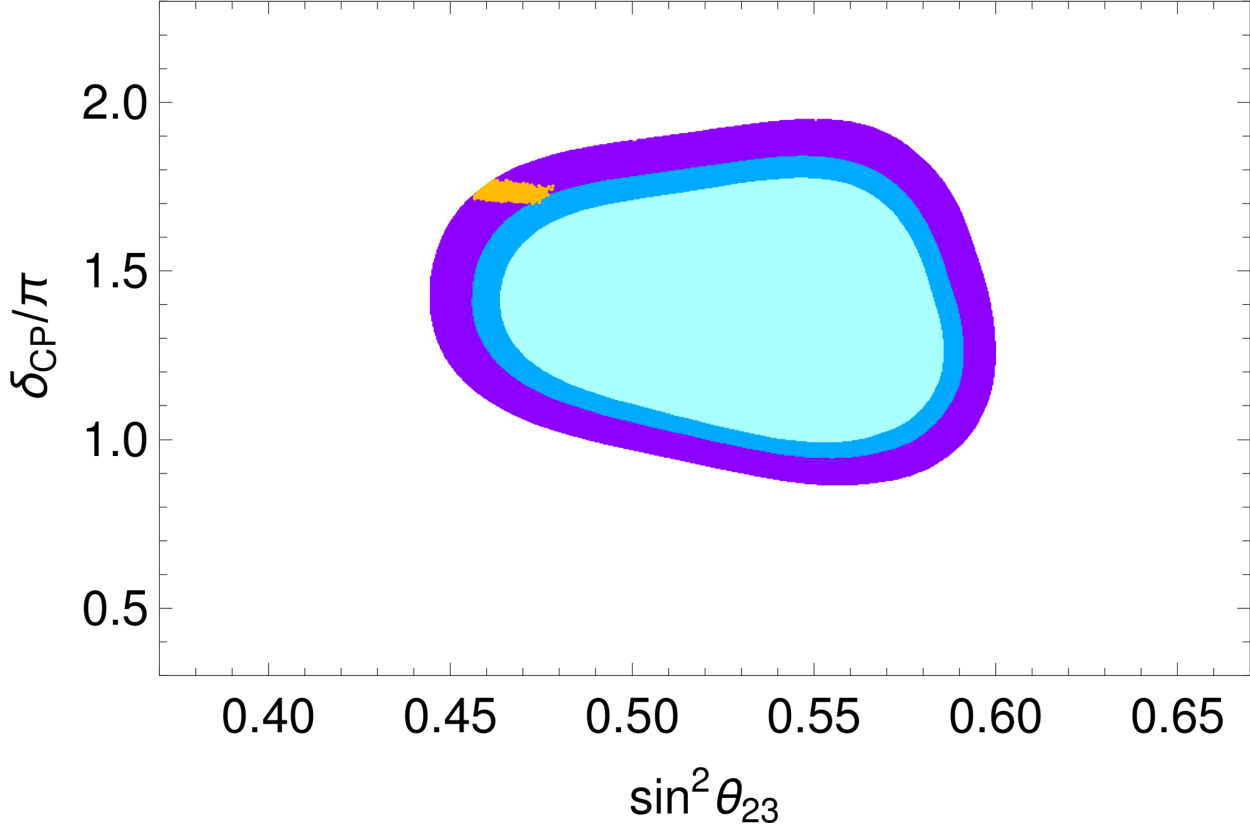}  }
\caption{Correlation between the observables predicted by the model: $\sin
^{2}\protect\theta _{12}$, $\sin ^{2}\protect\theta _{13}$, $\sin ^{2}%
\protect\theta _{23}$, $\protect\delta _{CP}$, $\Delta m_{21}^{2}$, and $%
\Delta m_{31}^{2}$ for the normal hierarchy, superimposed on the global fits from Ref.~\protect\cite%
{deSalas:2017kay}.
Model predictions are shown in orange, while the 90, 95,
and 99\% C.L. contours of the global fit are in purple, blue, and light blue, respectively.
}
\label{fig:corr}
\end{figure}

Tables \ref{tab:neutrinos-NH} and \ref{tab:neutrinos-IH} show the average model values and experimental values of the neutrino observables for both normal hierarchy (NH) and inverted hierarchy (IH).
Figure~\ref{fig:corr} shows several solutions consistent with the global fits for the NH (the IH has the same behavior).
The dots in orange correspond to the model values, which for comparison are plotted
over the experimental values taken from Ref.~\cite{deSalas:2017kay}.
To give an example, for each hierarchy we choose a representative value of the neutrino sector parameters: 
\begin{gather}
\begin{split}
a \simeq -0.474,\ \ b \simeq -0.367,\ \ c\simeq 0.487,\ \ d\simeq -0.0590\ \ \ \mbox{(NH)}\\
a \simeq -0.254,\ \ b \simeq 0.352,\ \ c\simeq -0.795,\ \ d\simeq 0.0174\ \ \ \mbox{(IH)}
\label{benchmark}
\end{split}
\end{gather}
which produces the following mass spectrum:
\begin{gather}
\begin{split}
m_1\simeq 5.84\mbox{ [meV]},\ \ m_2\simeq 10.5\mbox{ [meV]},\ \ m_3\simeq 50.4
\mbox{ [meV]}\ \ \ \mbox{(NH)}\\
m_1\simeq 50.4\mbox{ [meV]},\ \ m_2\simeq 51.1\mbox{ [meV]},\ \ m_3\simeq 11.7
\mbox{ [meV]}\ \ \ \mbox{(IH).}
\end{split}
\end{gather}
Thus, with four effective parameters, the model reproduces the experimental
values of the six physical observables of the neutrino sector, i.e., the
neutrino mass squared splittings $\Delta m_{21}^{2}$ and $\Delta m_{31}^{2}$%
, the leptonic mixing angles $\theta _{12}$, $\theta _{23}$, $\theta _{13}$,
and the leptonic Dirac $CP$ violating phase $\delta _{CP}$.
The model values are consistent with the current neutrino oscillation experimental data, for both normal and inverted mass ordering, as shown in Tables \ref{tab:neutrinos-NH} and \ref{tab:neutrinos-IH}.
From Figure~\ref{fig:corr}, we can see that for the normal hierarchy, $\Delta m_{31}^{2}$, $\Delta m_{21}^{2}$, $\sin
^{2}\theta _{12}$, and $\sin ^{2}\theta _{13}$ are evenly
distributed in the allowed range.
On the other hand, for $\sin ^{2}\theta
_{23}$ and $\delta _{CP}$, the model features more definite predictions.
The same behavior is found for the inverted hierarchy.

It is worth mentioning that in a generic scenario, the neutrino
Yukawa couplings are complex, thus the light active neutrino sector has eight
parameters. However, not all of them are physical. Considering the case of
real VEVs for the gauge-singlet scalars $\rho_1$, $\rho_2$, and $\xi$, the
phase redefinition of the leptonic fields $L_L$ and $N_R$ allows to rotate
away the phase of one of the neutrino Yukawa couplings, leading to seven
physical parameters. On the other hand, if we consider complex VEVs for the
gauge-singlet scalars $\rho_1$, $\rho_2$, and $\xi$, we can use their phases
to set three of the four neutrino Yukawa couplings real. Therefore, in this case we
are left with five effective parameters in the neutrino sector. However, for
the sake of simplicity, we are considering a particular benchmark scenario
with real neutrino Yukawa couplings, i.e., four effective parameters. In this
simplified benchmark scenario, the complex phase responsible for $CP$
violation in neutrino oscillation arises from the spontaneous breaking of
the $A_4$ discrete group. This mechanism for inducing $CP$ violation in the
fermion sector via the spontaneous breaking of discrete groups is the
so-called geometrical $CP$ violation \cite%
{Branco:1983tn,Chen:2011tj,Bhattacharyya:2012pi,Girardi:2013sza,Varzielas:2013sla,Chen:2014tpa,Branco:2015hea,CarcamoHernandez:2019eme,CarcamoHernandez:2019pmy,CarcamoHernandez:2019vih}%
.

Now we determine the effective Majorana neutrino mass parameter, which is
proportional to the neutrinoless double beta decay ($0\nu\beta\beta$)
amplitude. The effective Majorana neutrino mass parameter is given by
\begin{equation}
m_{\beta \beta }=\left\vert \sum_{j} m_{\nu_j} U_{ej}^2\right\vert,
\label{mee}
\end{equation}
where $U_{ej}^{2}$ and $m_{\nu_j}$ are the PMNS mixing matrix elements and
the Majorana neutrino masses, respectively. 
\begin{figure}[t]
\subfloat{\includegraphics[width=0.45\textwidth]{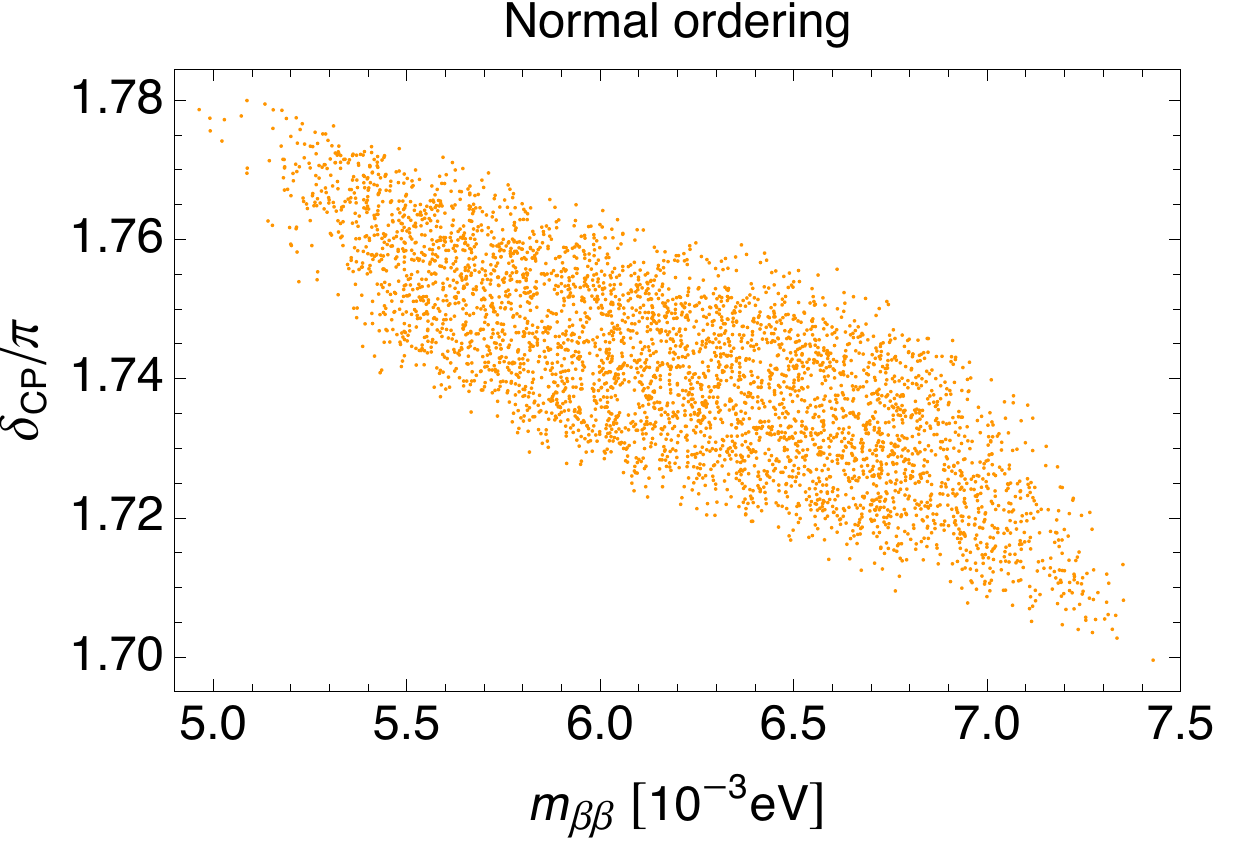}} \subfloat{  \ \ \
\ \includegraphics[width=0.45\textwidth]{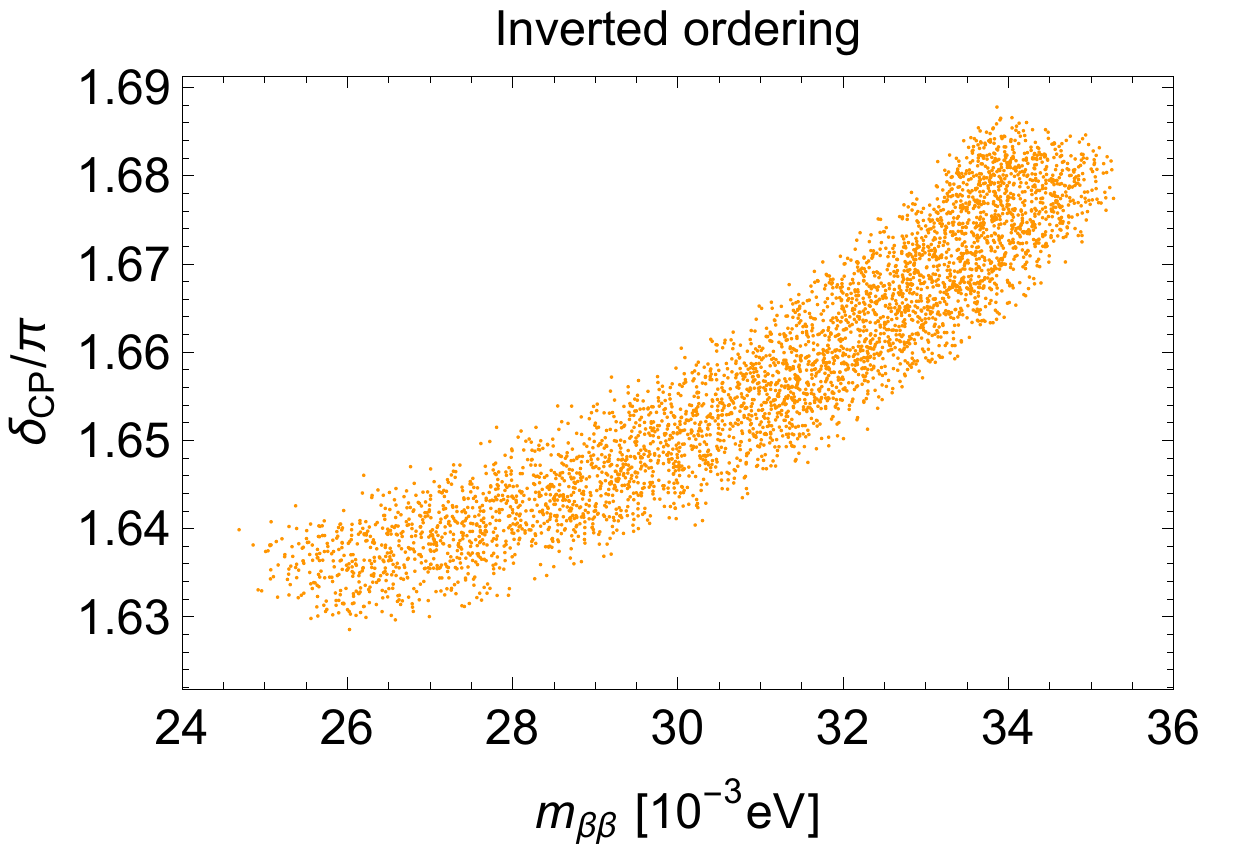}}
\caption{Model predictions for the Dirac $CP$ violating phase versus the effective Majorana mass parameter.}
\label{fig:mbb}
\end{figure}
As we can see from Figure~\ref{fig:mbb}, the predicted effective
Majorana neutrino mass parameter is within the range: 
\begin{equation}
m_{\beta \beta }= (5.0-7.4) \mbox{ meV} \mbox{ (NH)}, \ \ \ 
m_{\beta \beta }= (24.7-35.3) \mbox{ meV} \mbox{ (IH)} ,
\label{eff-mass-pred}
\end{equation}
which is below the sensitivity of present $0\nu \beta \beta $-decay
experiments. The current experimental sensitivity on the Majorana neutrino
mass parameter is obtained from the KamLAND-Zen limit on the ${}^{136}%
\mathrm{Xe}$ $0\nu\beta\beta$ decay half-life, $T^{0\nu\beta\beta}_{1/2}%
\left({}^{136}\mathrm{Xe}\right)\geq 1.07\times 10^{26}$ yr \cite%
{KamLAND-Zen:2016pfg}, which yields the corresponding upper limit on the
Majorana mass, $|m_{\beta\beta}|\leq (61-165)$ meV at 90\% C.L. For other $%
0\nu\beta\beta$-decay experiments see Refs.~\cite{Aalseth:2017btx,Alduino:2017ehq,Albert:2017owj,Alduino:2017pni,Albert:2014fya,Arnold:2016bed}%
. The experimental sensitivity of neutrinoless double beta decay searches is
expected to improve in the near future. Note that the model predicts a range
of values for neutrinoless double beta decay rates that can be tested by the
next-generation bolometric CUORE experiment \cite{Alduino:2017pni}, as well
as the next-to-next-generation ton-scale $0\nu \beta \beta $-decay
experiments \cite%
{KamLAND-Zen:2016pfg,Albert:2017owj,Abt:2004yk,Gilliss:2018lke}.

Finally, we briefly comment on the prospects of observing heavy neutrinos
with masses around $50\mbox{ GeV}$ in collider experiments. In the type-I
seesaw model, the heavy light mixing squared, $|U|^2$ is given by 
\begin{equation}
|U|^2 \sim \left(\frac{M_D}{M_N}\right)^2 \sim \frac{m_\nu}{m_N},
\end{equation}
which in our case (for $m_N \approx 50\mbox{ GeV}$ and $m_\nu \approx 50%
\mbox{ meV}$) gives $|U|^2 \sim O(10^{-12})$. Even though this is a very
small mixing, typical of the type-I seesaw model, for masses \mbox{$m_N\sim
50\mbox{ GeV}$} it might be within the reach of future colliders such as
the FCC-ee \cite{Blondel:2014bra}. The most sensitive channel at the FCC-ee
would be $Z\rightarrow \nu N$, when the decays of $N$ are fully
reconstructible, i.e., $N\rightarrow \ell W^* \rightarrow \ell q \bar
q^{\prime }$. According to the analysis in Ref. \cite{Blondel:2014bra}, most
of the backgrounds for this decay can be reduced if one takes into
consideration (i) the displaced vertex topology produced by the long-lived 
$N$ (expected for these small couplings) and (ii) the full reconstruction
of the heavy neutrino mass, allowed by its visible decay. For $\sim 10^{13}$ $%
Z$ decays, this would allow reaching sensitivities down to a heavy-light
mixing $|U|^2 \sim 10^{-12}$, for heavy neutrino masses between $40$ and $80%
\mbox{ GeV}$.

\section{Summary and Conclusions}

\label{conclusions}

We have proposed a low scale seesaw model based on the $A_{4}{\times Z}_{4}$
discrete symmetry, where the masses for the light active neutrinos are
produced by a type-I seesaw mechanism mediated by three $\sim 50$ GeV scale
right-handed Majorana neutrinos. Contrary to the original type I seesaw,
where the right-handed neutrinos are required to have masses much larger
than the electroweak scale to reproduce the light active neutrino mass scale 
$m_\nu$, in this case $m_\nu$ is suppressed by the ratio between the
discrete symmetry breaking scale ($v_S$) and the cutoff ($\Lambda$) of the
model: $m_\nu \propto \left(v_S/\Lambda \right)^2$. That is, the large $%
\Lambda/v_S$ ratio plays the role of the heavy mass scale in the original
seesaw. This allows lighter Majorana masses, that might be eventually tested
at future colliders such as the FCC-ee.

The model is predictive in the sense that it reproduces the experimental
values of the six low energy neutrino observables with only four effective
parameters. Two of the predicted observables (being within the 90\% C.L.
global-fit regions) are not aligned with the central values of the global fits, so are
distinctive predictions of the model. These are the $CP$-violating angle,
predicted to be
\begin{equation}
\delta_{CP}=
\begin{cases}
312.9^\circ \pm 2.4^\circ, \text{   (NH)}\\
297.2^\circ \pm 2.7^\circ, \text{   (IH),}
\end{cases}
\end{equation}
and the ``atmospheric'' neutrino mixing angle
\begin{equation}
\sin^2\theta_{23}=
\begin{cases}
0.465 \pm 0.004, \text{   (NH)}\\
0.565\pm 0.001, \text{   (IH).}
\end{cases}
\end{equation}
The effective Majorana neutrino mass parameter is predicted to be
\begin{equation}
m_{\beta\beta}=
\begin{cases}
(6.2\pm 0.5)\mbox{ meV}, \text{   (NH)}\\
(31.1\pm 2.6)\mbox{ meV}, \text{   (IH).}
\end{cases}
\end{equation}
The scalar sector of the model corresponds to the SM Higgs doublet
supplemented with additional singlet scalars. The phenomenology of this kind
of extended Higgs sectors is well studied, and many direct and indirect
searches have been proposed in the literature. For masses of the additional
scalar singlets ($m_S$) in the range $1\mbox{ TeV} \lesssim m_S \lesssim 11%
\mbox{ TeV}$, the scalar sector would be within the reach of future
colliders.

\section*{Acknowledgments}

This work was supported by FONDECYT (Chile) under Grants No.~1170803,
No.~3170906 and No.~11180873, and in part by Conicyt PIA/Basal FB0821.

\appendix

\section{The product rules for $A_4$}

\label{ap}

The $A_{4}$ group has one three-dimensional $\mathbf{3}$\ and three distinct
one-dimensional $\mathbf{1}$, $\mathbf{1}^{\prime }$, and $\mathbf{1}^{\prime
\prime }$ irreducible representations, satisfying the following product
rules: 
\begin{eqnarray}
&&\hspace{18mm}\mathbf{3}\otimes \mathbf{3}=\mathbf{3}_{s}\oplus \mathbf{3}%
_{a}\oplus \mathbf{1}\oplus \mathbf{1}^{\prime }\oplus \mathbf{1}^{\prime
\prime },  \label{A4-singlet-multiplication} \\[0.12in]
&&\mathbf{1}\otimes \mathbf{1}=\mathbf{1},\hspace{5mm}\mathbf{1}^{\prime
}\otimes \mathbf{1}^{\prime \prime }=\mathbf{1},\hspace{5mm}\mathbf{1}%
^{\prime }\otimes \mathbf{1}^{\prime }=\mathbf{1}^{\prime \prime },\hspace{%
5mm}\mathbf{1}^{\prime \prime }\otimes \mathbf{1}^{\prime \prime }=\mathbf{1}%
^{\prime }.  \notag
\end{eqnarray}%
Considering $\left( x_{1},y_{1},z_{1}\right) $ and $\left(
x_{2},y_{2},z_{2}\right) $ as the basis vectors for two $A_{4}$-triplets $%
\mathbf{3}$, the following relations are fulfilled: 
\begin{eqnarray}
&&\left( \mathbf{3}\otimes \mathbf{3}\right) _{\mathbf{1}%
}=x_{1}y_{1}+x_{2}y_{2}+x_{3}y_{3},  \label{triplet-vectors} \\
&&\left( \mathbf{3}\otimes \mathbf{3}\right) _{\mathbf{3}_{s}}=\left(
x_{2}y_{3}+x_{3}y_{2},x_{3}y_{1}+x_{1}y_{3},x_{1}y_{2}+x_{2}y_{1}\right) ,\
\ \ \ \left( \mathbf{3}\otimes \mathbf{3}\right) _{\mathbf{1}^{\prime
}}=x_{1}y_{1}+\omega x_{2}y_{2}+\omega ^{2}x_{3}y_{3},  \notag \\
&&\left( \mathbf{3}\otimes \mathbf{3}\right) _{\mathbf{3}_{a}}=\left(
x_{2}y_{3}-x_{3}y_{2},x_{3}y_{1}-x_{1}y_{3},x_{1}y_{2}-x_{2}y_{1}\right) ,\
\ \ \left( \mathbf{3}\otimes \mathbf{3}\right) _{\mathbf{1}^{\prime \prime
}}=x_{1}y_{1}+\omega ^{2}x_{2}y_{2}+\omega x_{3}y_{3},  \notag
\end{eqnarray}
where $\omega =e^{i\frac{2\pi }{3}}$. The representation $\mathbf{1}$ is
trivial, while the nontrivial $\mathbf{1}^{\prime }$ and $\mathbf{1}%
^{\prime \prime }$ are complex conjugate to each other. Reviews of discrete
symmetries in particle physics can be found in Refs. \cite%
{Ishimori:2010au,Altarelli:2010gt,King:2013eh, King:2014nza}.

\end{document}